\documentclass[twocolumn,showpacs,preprintnumbers,amsmath,amssymb,widetext]{revtex4}
\usepackage{graphicx}
\begin{document}
\title {Josephson tunnel junction controlled by quasiparticle injection}

\author{Francesco Giazotto}
\email{giazotto@sns.it}
\affiliation{NEST-INFM and Scuola Normale Superiore, I-56126 Pisa, Italy}
\author{Jukka P. Pekola}
\affiliation{Low Temperature Laboratory, Helsinki University of Technology, P.O. Box 3500, 
FIN-02015 HUT, Finland}

\begin{abstract}
A Josephson tunnel junction transistor based on quasiparticle injection is proposed. Its operation 
relies on the manipulation of the electron distribution in one of the junction  electrodes. 
This is accomplished by injecting quasiparticle current through the junction electrode by two 
additional tunnel coupled superconductors. Both large supercurrent enhancement and fast quenching 
can be achieved with respect to equilibrium by varying quasiparticle injection for proper 
temperature regimes and suitable superconductor combinations. Joined with large power 
gain this makes the device attractive for applications where reduced noise and low power dissipation are 
required.

\end{abstract}


\maketitle

The control of Josephson currents as for the realization of efficient transistors has gained 
recently  a rekindled interest \cite{golubov}. 
A novel development in mesoscopic superconductivity is indeed represented by controllable 
superconductor(S)-normal metal(N)-superconductor(S) metallic weak links \cite{wilhelm}, where 
supercurrent suppression is achieved by altering the quasiparticle distribution in the N region 
through current injection. So far there have been a few successful demonstrations of this 
operation principle \cite{sns}. On the other hand, as  recently proposed \cite{sinis} and 
experimentally demonstrated \cite{transistor}, a SINIS control line (where I is a tunnel barrier) 
is particularly suitable for tuning Josephson current, allowing both enhancement and suppression 
with respect to equilibrium. Operation of these devices is based on the modification of the 
quasiparticle distribution in the N region of the junction.
In this letter, we propose an \emph{all}-superconducting tunnel junction device  in which 
transistor effect is obtained by driving the electron distribution out of equilibrium in the 
superconductor. This is performed by voltage biasing  a SISIS line (see Fig. 1) where the 
interelectrode is one of the two terminals belonging to the Josephson junction. 

As compared to the hybrid devices above the present one benefits from the sharp characteristics due to 
the presence of superconductors with unequal energy gaps. 
We consider different superconductors S$_1$ and S$_2$ with energy gaps $\Delta_1$ and $\Delta_2$ 
(and critical temperatures $T_{c1,2}$), respectively, and we assume $\Delta_2 <\Delta_1$ 
\cite{frank,manninen}. Under voltage bias $V_C$ across the S$_1$IS$_2$IS$_1$ line (see the inset 
of Fig. 1) the heat current from S$_2$ to S$_1$ is given by
\begin{equation}
\mathcal{P}=\frac{2}{e^2 R_T}\int _{-\infty} ^\infty d\varepsilon \varepsilon 
\mathcal{N}_1(\tilde{\varepsilon})\mathcal{N}_2(\varepsilon)[f_0(\varepsilon,T_{e2})-
f_0(\tilde{\varepsilon},T_{e1})],
\end{equation}
where $\tilde{\varepsilon}=\varepsilon-eV_C/2$, $f_0(\varepsilon,T)$ is the Fermi-Dirac 
distribution function, $T_{ek}$ is the electron temperature in S$_k$, $R_T$ is the normal-state 
resistance of each S$_1$IS$_2$ junction and 
$\mathcal{N}_k(\varepsilon)=|\mathrm{Re}[(\varepsilon+i\Gamma_k)/\sqrt{(\varepsilon+i\Gamma_k)^2-\Delta_k^2}]|$ is the smeared \cite{cooler} BCS density of states of S$_k$. Figure 1 shows the 
calculated \cite{gamma} heat current  versus bias voltage $V_C$ at constant bath temperature 
$T_{bath}=T_{e1}=T_{e2}=0.4\,T_{c1}$ and for different values of $\Delta_2$. 
\begin{figure}[ht]
\begin{center}
\includegraphics[width=7.5cm,clip]{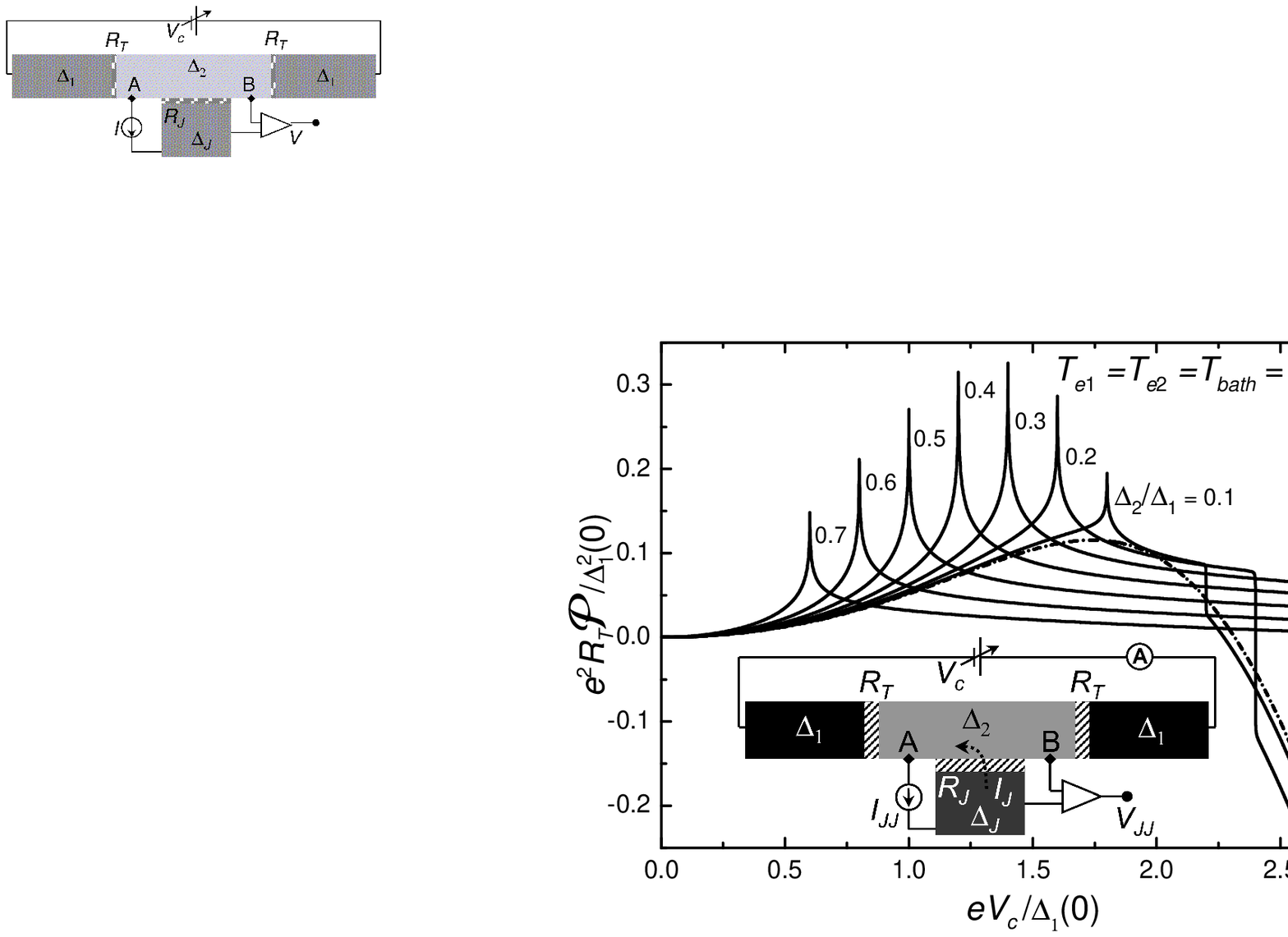}
\end{center}
\caption{Heat current $\mathcal{P}$ out from S$_2$ by a S$_1$IS$_2$IS$_1$ line  vs control voltage 
$V_C$ at $T_{e1}=T_{e2}=T_{bath}=0.4\,T_{c1}$ for several $\Delta_2/\Delta_1$ ratios. Dash-dotted 
line represents $\mathcal{P}$ when S$_2$ is in the normal state. Inset: Scheme of the  Josephson 
device. The bias $V_C$ across the S$_1$IS$_2$IS$_1$ line allows to control the supercurrent $I_J$ 
(along the dashed line) increasing or suppressing its amplitude with respect to equilibrium. 
\textbf{A} and \textbf{B} represent tunnel contacts used to inject and measure the supercurrent.} 
\label{device}
\end{figure}
$\mathcal{P}$ is  symmetric  in $V_C$ and it is positive for $V_C<2|\Delta_1(T)+\Delta_2(T)|/e$ 
thus allowing heat removal from S$_2$, i.e., \textit{hot} quasiparticle excitations are 
transferred to S$_1$; furthermore, the heat current is maximized at  
$V_C=\pm2|\Delta_1(T)-\Delta_2(T)|/e$, where the finite-temperature logarithmic singularity occurs
 \cite{frank} (in a real situation it will be somewhat broadened 
by smearing in the density of states \cite{frank,manninen,cooler}). 
From  Fig. 1  it follows that a positive heat current from S$_2$ exists only if 
$\Delta_2(T)<\Delta_1(T)$ holds.  The dash-dotted line represents the heat current in the system 
when S$_2$ is in the normal state. Notably,  when S$_2$ is in the superconducting state 
$\mathcal{P}$ can largely exceed that one in the normal state. Then, on approaching 
$V_C=\pm2|\Delta_1(T)+\Delta_2(T)|/e$, a sharp transition brings $\mathcal{P}$ to negative 
values. 
An additional superconducting electrode S$_J$ is connected to S$_2$ through a tunnel barrier so to 
realize a S$_J$IS$_2$ Josephson junction. S$_J$ is characterized by its own energy gap $\Delta_J$ 
(different in general from $\Delta_{1,2}$) with critical temperature $T_{cJ}$, and $R_J$ is the  
normal-state resistance of the junction. As we shall prove this transistor operation relies on the 
quasiparticle distribution established in S$_2$ upon voltage biasing the control line.

We consider a transport regime where strong inelastic electron-electron interaction forces the  
system to retain a local thermal (\emph{quasi})equilibrium, so that the quasiparticle distribution 
in S$_2$ is described by a Fermi function at  temperature $T_{e2}$ differing in general from 
$T_{bath}$. In order to determine the actual $T_{e2}$ upon biasing with $V_C$ we need to include 
those scattering mechanisms that transfer energy in S$_2$. 
At the typical operation temperatures the predominant contribution comes from electron-phonon scattering that transfers energy 
between electrons and phonons. This heat flux is given by 
$\mathcal{P}_{e2-bath}=\Sigma\mathcal{V}(T_{e2}^5-T_{bath}^5)$ \cite{urbina}, where $\Sigma$ is a 
material-dependent parameter and $\mathcal{V}$ is the volume of S$_2$.  The temperature $T_{e2}$ 
is then determined by solving the energy-balance equation 
$\mathcal{P}(V_C,T_{bath},T_{e2})+\mathcal{P}_{e2-bath}=0$.

The supercurrent ($I_J$) flowing through the S$_J$IS$_2$ junction can be calculated from 
\cite{golubov,tero}:
\begin{equation}
\begin{split}
I_J=-\frac{\sin\phi}{2eR_J}\int_{-\infty}^\infty
d\varepsilon\{\mathbf{f}_2(\varepsilon)\mathrm{Re}\mathcal{F}_2(\varepsilon)\mathrm{Im}\mathcal{F}
_J(\varepsilon)+
\\
+\mathbf{f}_J(\varepsilon)\mathrm{Re}\mathcal{F}_J(\varepsilon)\mathrm{Im}\mathcal{F}_2(\varepsilon)\},\,\,\,\,\,
\end{split}
\end{equation}
where $\phi$ is the phase difference between the superconductors, 
$\mathbf{f}_{2,J}(\varepsilon)=\tanh[\varepsilon/2k_BT_{e2,bath}]$ and 
$\mathcal{F}_{2,J}(\varepsilon)=\Delta_{2,J}/\sqrt{(\varepsilon+i\Gamma_{2,J})^2-\Delta_{2,J}^2}$. 
In the aforementioned expressions we set $\Delta_{2}=\Delta_{2}(T_{e2})$ and 
$\Delta_{J}=\Delta_{J}(T_{bath})$.  Equation (2) shows that, for fixed $T_{bath}$ and phase 
difference, the Josephson current is controlled by $T_{e2}$.
\begin{figure}[ht]
\begin{center}
\includegraphics[width=7.0cm,clip]{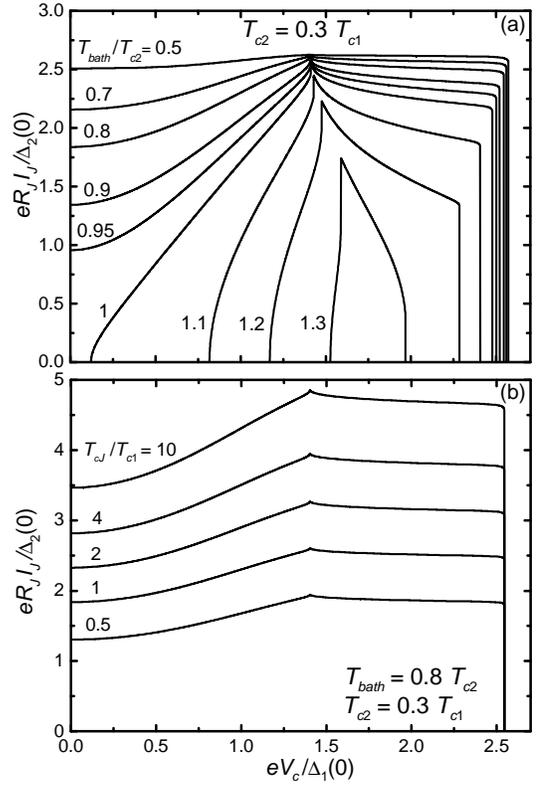}
\end{center}
\caption{(a) Supercurrent $I_J$ vs control voltage $V_C$ calculated at different bath temperatures 
$T_{bath}$ for $T_{c2}=0.3\,T_{c1}$ (corresponding roughly to the Ti/Al combination) and 
$T_{cJ}=T_{c1}$. Note the sharp $I_J$ suppression at 
$eV_C=2[\Delta_1(T_{bath})+\Delta_2(T_{e2})]$. (b) Supercurrent vs $V_C$ calculated for several 
$T_{cJ}/T_{c1}$ ratios at $T_{bath}=0.8\,T_{c2}$ and for $T_{c2}=0.3\,T_{c1}$.}
\label{supercurrentbehavior}
\end{figure}
The solution of the balance equation for $T_{e2}$ combined with Eq. (2) yields the dimensionless 
transistor output characteristic shown in Fig. 2(a) \cite{parameters}, where $I_J$ is plotted 
versus  $V_C$ at different bath temperatures, for $T_{c2}=0.3\,T_{c1}$ and $T_{cJ}=T_{c1}$.  For 
$T_{bath}<T_{c2}$, $I_J$ first increases monotonically up to 
$eV_C=2[\Delta_1(T_{bath})-\Delta_2(T_{e2})]$, where the cooling power is maximized; then it 
starts to slightly decrease after which it is rapidly quenched at 
$eV_C=2[\Delta_1(T_{bath})+\Delta_2(T_{e2})]$. Notably, even at bath temperatures exceeding 
$T_{c2}$ (i.e., for $T_{bath}\geq T_{c2}$ where $I_J$ is zero at equilibrium), a finite 
supercurrent is obtained at a voltage for which S$_2$ is brought into the superconducting state, 
after which $I_J$
is  recovered up to a large extent. The influence of different S$_J$ on the  supercurrent is 
displayed in Fig. 2(b) that shows  $I_J$ versus $V_C$ at $T_{bath}=0.8\,T_{c2}$ for different 
$T_{cJ}/T_{c1}$ ratios. As a consequence $I_J$ is  enhanced upon increasing $\Delta_J$ , being 
nearly doubled  for $T_{cJ}/T_{c1}=10$. 

\begin{figure}[ht]
\begin{center}
\includegraphics[width=7.5cm,clip]{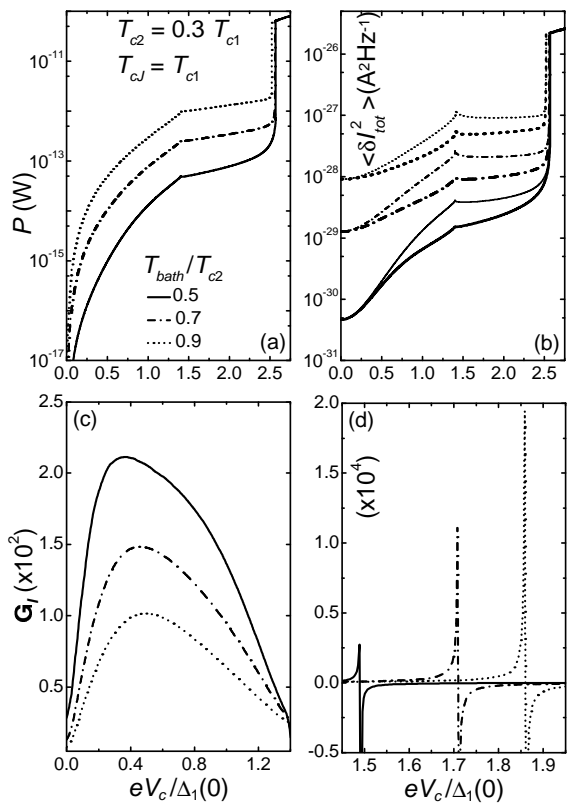}
\end{center}
\caption{(a)  Dissipated power $P$ and (b) total input noise $\langle \delta I_{tot}^2 \rangle$ in the 
S$_1$IS$_2$IS$_1$ line against $V_C$. The transistor current gain $\textsf{G}_I(V_C)$ is shown in 
(c) and (d) in two different ranges of $V_C$. All these calculations are performed for 
$T_{c2}=0.3\,T_{c1}$, $T_{cJ}=T_{c1}$ and at three different bath temperatures.}
\label{supercurrentbehavior}
\end{figure}

Figure 3(a) displays the transistor power dissipation $P=V_CI_C$, where $I_C$ is the control 
current in the S$_1$IS$_2$IS$_1$ line, calculated for $T_{c2}=0.3\,T_{c1}$ and  $T_{cJ}=\,T_{c1}$ 
at different bath temperatures. The plot shows that at the lowest temperatures $P$ obtains values 
of the order of some fW in the regime of supercurrent enhancement while of some hundred of fW 
around the $I_J$ quenching. This is because of  low control currents through the structure. As far 
as noise is concerned, the total input noise per unit bandwidth $\langle \delta I_{tot}^2 \rangle$ 
\cite{noise} in the control line can be expressed as \cite{golubev}
\begin{equation}
\langle \delta I_{tot}^2 \rangle=\langle\delta I_C^2\rangle -2S_{I_C}\langle\delta 
\mathcal{P}\delta I_{C}\rangle+S_{I_C}^2\langle\delta \mathcal{P}^2\rangle,
\end{equation} 
where 
\begin{equation}
\langle\delta I_C^2\rangle=\frac{1}{R_T}\int _{-\infty}^\infty d\varepsilon 
\mathcal{N}_1(\tilde{\varepsilon})\mathcal{N}_2(\varepsilon)\mathcal{W}(\varepsilon,\tilde{\varepsilon}),
\end{equation}
\begin{equation}
\langle\delta \mathcal{P}^2\rangle=\frac{1}{e^2 R_T}\int _{-\infty}^\infty d\varepsilon 
\varepsilon^2 
\mathcal{N}_1(\tilde{\varepsilon})\mathcal{N}_2(\varepsilon)\mathcal{W}(\varepsilon,\tilde{\varepsilon}),
\end{equation}
\begin{equation}
\langle\delta \mathcal{P}\delta I_{C} \rangle=-\frac{1}{e R_T}\int _{-\infty}^\infty d\varepsilon 
\varepsilon 
\mathcal{N}_1(\tilde{\varepsilon})\mathcal{N}_2(\varepsilon)\mathcal{W}(\varepsilon,\tilde{\varepsilon}),
\end{equation}
and 
$\mathcal{W}(\varepsilon,\tilde{\varepsilon})=f_0(\varepsilon,T_{e2})(1-f_0(\tilde{\varepsilon},T_
{bath}))
+f_0(\tilde{\varepsilon},T_{bath})(1-f_0(\varepsilon,T_{e2}))$.
Equations (4), (5) and (6) represent fluctuations due to charge and heat flow, and their mutual 
correlation, respectively, and $S_{I_C}$ is the zero-frequency current responsivity,  
$S_{I_C}(V_C)=(\partial I_C/\partial T_{e2})/(5\Sigma\mathcal{V}T_{e2}^4+\partial 
\mathcal{P}/\partial T_{e2})$ \cite{golubev}. $\langle \delta I_{tot}^2 \rangle$ is displayed in 
Fig. 3(b) for the same parameters as in Fig. 3(a), and shows that input noise as low as 
some $10^{-30}$ A$^2$ Hz$^{-1}$ can be achieved in the enhancement regime while of some $10^{-29}$ 
A$^2$ Hz$^{-1}$ at the quenching voltage. Thin lines are the \textit{uncorrelated} noise power, 
i.e., the noise obtained by adding the contributions of Eqs. (4) and (5) only. Notably, the impact of 
mutual correlations (Eq. (6)) is easily recognized leading to significant noise 
reduction ($\sim 50\%$)  in the range of supercurrent enhancement.

We shall further comment on the available gain. Input ($V_{in}=V_C\sim \Delta_1$) and output 
($V_{out}=I_JR_J\sim \Delta_2$) (see also Fig. 2(b)) voltages allow a voltage gain 
$\textsf{G}_V=V_{out}/V_{in}\sim \Delta_2/\Delta_1$ so that with realistic parameters 
$\textsf{G}_V$ is not much smaller than 1. The differential current gain, defined as 
$\textsf{G}_I=dI_J/dI_C=(\frac{dI_J}{dV_C})(\frac{dI_C}{dV_C})^{-1}$, is plotted in Fig. 3(c,d) in 
two different bias  ranges for some values of $T_{bath}$. The figure shows that $\textsf{G}_I$ 
obtains large values with some $10^2$ in the regime of supercurrent enhancement and several $10^3$ 
below the quenching. The corresponding input impedance ranges from hundreds of k$\Omega$ to tens 
of M$\Omega$, respectively. 
In order to exploit the power gain ($\textsf{G}_P$) the Josephson junction needs to be operated in 
the dissipative regime; in such a situation an estimate for the achievable power gain \cite{sinis} 
yields $\textsf{G}_P\sim 10^2 ... 10^3$ depending on the operating $V_C$ and bias current $I_{JJ}$ 
across the junction (see Fig. 1). One should note that such a large power gain, not achievable, 
e.g., using a SINIS controlled SNS transistor \cite{sinis} in the same transport regime, is an additional advantage of the 
present scheme. 

We conclude with some further benefits of our proposal. Due to the presence of the superconducting
interelectrode, highly transmissive tunnel junctions are not necessary unlike in SINIS devices.
The device is also less sensitive to thermal fluctuations as compared to SNS junctions 
\cite{transistor}. Furthermore, it is easier to fabricate taking advantage of the well established metal-based tunnel junction technology.
A promising choice for transistor 
and  switch implementations could be a combination of Al and Ti.

The authors acknowledge the Large Scale Installation Program ULTI-3 of the European Union and the Academy of Finland (TULE program) for financial support, and D. Golubev, T. T. Heikkil\"{a}, A. M. Savin and H. Sepp\"{a} for fruitful discussions.



\begin{references}

\bibitem{golubov}
    See, for example, A.A. Golubov, M.Yu. Kupriyanov, and E. Il'ichev, Rev. Mod. Phys. 
\textbf{76}, 411 (2004).


\bibitem{wilhelm}
    F.K. Wilhelm, G. Sch\"{o}n, and A.D. Zaikin, Phys. Rev. Lett. \textbf{81}, 1682 (1998).


\bibitem{sns}
    J.J.A. Baselmans, A.F. Morpurgo, B.J. van Wees, and T.M. Klapwijk, Nature (London) {\bf 397}, 
43 (1999);
    J. Huang, F. Pierre, T.T. Heikkil\"{a}, F.K. Wilhelm, and N.O. Birge, Phys. Rev. B {\bf 66}, 
020507 (2002);
    R. Shaikhaidarov, A.F. Volkov, H. Takayanagi, V.T. Petrashov, and P. Delsing, Phys. Rev. B 
{\bf 62}, R14649       (2000).


\bibitem{sinis}
    F. Giazotto, T.T. Heikkil\"{a}, F. Taddei, R. Fazio, J.P. Pekola, and F. Beltram, Phys. Rev. 
Lett. \textbf{92}, 137001 (2004); F. Giazotto, F. Taddei, T.T. Heikkil\"{a}, R. Fazio, and F. 
Beltram, Appl. Phys. Lett. \textbf{83}, 2877 (2003).


\bibitem{transistor}
    A.M. Savin, J.P. Pekola, J.T. Flyktman, A. Anthore, and F. Giazotto, Appl. Phys. Lett. 
\textbf{84}, 4179 (2004).

\bibitem{frank}
    B. Frank and W. Krech, Phys. Lett. A \textbf{235}, 281 (1997).


\bibitem{manninen}
    A.J. Manninen, J.K. Suoknuuti, M.M. Leivo, and J.P. Pekola, Appl. Phys. Lett. \textbf{74}, 
3020 (1999).


\bibitem{cooler}
    J.P. Pekola, T.T. Heikkil\"{a}, A.M. Savin, J.T. Flyktman, F. Giazotto, and F.W.J. Hekking, 
Phys. Rev. Lett. \textbf{92}, 056804 (2004).



\bibitem{gamma}
    We assume throughout the paper a realistic smearing parameter $\Gamma_k=10^{-4}\Delta_k$ (see 
Ref. \cite{cooler}).


\bibitem{urbina}
    F.C. Wellstood, C. Urbina, and J. Clarke, Phys. Rev. B \textbf{49}, 5942 (1994).


\bibitem{tero} T.T. Heikkil\"a, private communication.


\bibitem{parameters}
    For the calculation we choose $\phi=\pi/2$, $T_{c1}=1.19$ K (Al), $R_T=10^3\,\Omega$, 
$R_J=300\,\Omega$, $\mathcal{V}=0.1\,\mu$m$^3$ and $\Sigma=10^{-9}$ WK$^{-5}$$\mu$m$^{-3}$ (Ti) 
(see Ref. \cite{manninen}).


\bibitem{noise}
    We suppose each junction to contribute in an uncorrelated way to the total noise.


\bibitem{golubev}
    D. Golubev and L. Kuzmin, J. Appl. Phys. \textbf{89}, 6464 (2001).





\end{references}
\end{document}